\documentclass[twocolumn,aps,longbibliography]{revtex4-1}

\pdfpageattr {/Group << /S /Transparency /I true /CS /DeviceRGB>>}

\usepackage{graphicx} % Include figure files
\usepackage{overpic}

\newcommand{\ket}[1]{\left|#1\right\rangle}

\DeclareSymbolFont{lettersA}{U}{txmia}{m}{it}
\DeclareMathSymbol{\real}{\mathord}{lettersA}{"92}
\DeclareMathSymbol{\field}{\mathord}{lettersA}{"83}

\begin{document}

\title{Topological code Autotune}

\author{Austin G. Fowler$^1$, Adam C. Whiteside$^1$, Angus L. McInnes$^1$, Alimohammad Rabbani$^2$}

\affiliation{$^1$Centre for Quantum Computation and Communication
Technology, School of Physics, The University of Melbourne, Victoria
3010, Australia,\\
$^2$Sharif University of Technology, Department of Computer Engineering, Tehran, Iran}

\date{\today}

\begin{abstract}
Many quantum systems are being investigated in the hope of building a large-scale quantum computer. All of these systems suffer from decoherence, resulting in errors during the execution of quantum gates. Quantum error correction enables reliable quantum computation given unreliable hardware. Unoptimized topological quantum error correction (TQEC), while still effective, performs very suboptimally, especially at low error rates. Hand optimizing the classical processing associated with a TQEC scheme for a specific system to achieve better error tolerance can be extremely laborious. We describe a tool Autotune capable of performing this optimization automatically, and give two highly distinct examples of its use and extreme outperformance of unoptimized TQEC. Autotune is designed to facilitate the precise study of real hardware running TQEC with every quantum gate having a realistic, physics-based error model.
\end{abstract}

\maketitle

Many quantum algorithms now exist, including factoring \cite{Shor94b}, searching \cite{Grov96}, simulating quantum physics \cite{Lloy96}, problems in knot theory \cite{Subr02}, and much more \cite{Jord10}. Large-scale simulations of topological quantum error correction (TQEC) indicate that gate error rates between 0.2\% and 0.5\% are sufficiently low to enable practical overhead, high reliability quantum computation \cite{Wang10,Fowl11b}. This is tantalisingly close to experimentally achieved two-qubit gate error rates of 2\% \cite{Monz11}, the best achieved to date in a system with the potential to implement the required 2-D array of qubits. This motivates the serious study of mapping TQEC schemes to physical hardware to enable realistic engineering trade-offs to be determined and optimizations found. We would very much like to collaborate with any experimentalist with a potentially 2-D qubit system and an interest in making use of TQEC. Given the transversely invariant nature of TQEC, experiments involving as few as two qubits can be sufficient to determine whether a system could successfully implement TQEC (see Appendix~\ref{2q_expt}).

Neutral atoms in optical lattices \cite{Jaks04} motivated the development of TQEC. Optical lattices lack the ability to easily implement arbitrary patterns of two-qubit gates, making them unsuitable for other types of quantum error correction \cite{Shor95,Cald95,Stea96,Knil04c,Baco06}. 2-D architectures designed for TQEC have since been developed for phosphorus atoms in silicon \cite{Holl06}, nitrogen-vacancy color centers in diamond \cite{Devi08}, superconducting circuits \cite{DiVi09}, quantum dots probabilistically entangled using linear optics \cite{Herr10}, quantum dots deterministically entangled using nonlinear optics \cite{Jone10}, and ion traps \cite{Stoc08,Monr12}. Basic TQEC has been experimentally demonstrated using linear optics \cite{Yao12}. In short, the best architectures in all scalable quantum computer technologies now make use of TQEC.

Every effort has been made to make this paper self-contained. Required quantum information background is provided in Section~\ref{qinf}. In Section~\ref{TQEC}, TQEC is defined and examples given. Automated methods of analyzing and visualizing the propagation of errors when using TQEC are presented in Section~\ref{Autotune}. The extreme performance difference of unoptimized TQEC and Autotuned TQEC is numerically demonstrated in Section~\ref{versus}. Section~\ref{Conclusion} concludes with a discussion of our planned future extensions of Autotune.

\section{Quantum information}
\label{qinf}

Quantum computers manipulate quantum systems with two relatively stable quantum states that are denoted $\ket{0}$ and $\ket{1}$. These quantum systems are called qubits. Unlike classical bits, which can be either 0 or 1, qubits can be placed in arbitrary superpositions $\ket{\Psi}=\alpha\ket{0}+\beta\ket{1}$, where $\alpha,\beta\in\field$ and $|\alpha|^2 + |\beta|^2=1$. The quantities $|\alpha|^2$ and $|\beta|^2$ represent the probabilities that the qubit, if measured, will be observed to be $\ket{0}$ or $\ket{1}$, respectively.

In addition to initialization to $\ket{0}$ and measurement, we will initially be interested in two quantum gates, Hadamard and controlled-NOT. To define their action we first define
\begin{eqnarray}
\ket{0} & = & \left( \begin{array}{c} 1 \\ 0 \end{array} \right), \\
\ket{1} & = & \left( \begin{array}{c} 0 \\ 1 \end{array} \right),
\end{eqnarray}
leading to
\begin{eqnarray}
\ket{\Psi}=\alpha\ket{0}+\beta\ket{1} & = & \left( \begin{array}{c} \alpha \\ \beta \end{array} \right).
\end{eqnarray}
Given two qubits in states $\ket{\Psi_1}=\alpha_1\ket{0}+\beta_1\ket{1}$ and $\ket{\Psi_2}=\alpha_2\ket{0}+\beta_2\ket{1}$, the state $\ket{\Psi_1}\ket{\Psi_2}$ corresponds to the outer product
\begin{eqnarray}
\ket{\Psi_1}\otimes\ket{\Psi_2} & = & \left( \begin{array}{c} \alpha_1\alpha_2 \\ \alpha_1\beta_2 \\ \beta_1\alpha_2 \\ \beta_1\beta_2 \end{array} \right).
\end{eqnarray}

Given the above definitions, Hadamard ($H$) is a single-qubit gate
\begin{eqnarray}
H & = & \frac{1}{\sqrt{2}}\left( \begin{array}{cc} 1 & 1 \\ 1 & -1 \end{array} \right),
\end{eqnarray}
and controlled-NOT (CNOT or $C_X$) is a two-qubit gate
\begin{eqnarray}
C_X & = & \left( \begin{array}{cccc} 1 & 0 & 0 & 0 \\ 0 & 1 & 0 & 0 \\ 0 & 0 & 0 & 1 \\ 0 & 0 & 1 & 0 \end{array} \right).
\end{eqnarray}
Note that $H$ and CNOT are unitary matrices. This is a general property of quantum gates other than initialization and measurement. In this instance, $H$ and CNOT are also Hermitian, and therefore self inverses. As defined above, CNOT flips the value of the second (target) qubit if the value of the first (control) qubit is $\ket{1}$ and does nothing otherwise.

Very general quantum errors can be expressed in terms of only $X$ and $Z$ errors \cite{Shor95}, where
\begin{eqnarray}
X & = & \left( \begin{array}{cc} 0 & 1 \\ 1 & 0 \end{array} \right), \\
Z & = & \left( \begin{array}{cc} 1 & 0 \\ 0 & -1 \end{array} \right).
\end{eqnarray}
A state $\ket{\Psi}$ that contains an $X$ error and is then acted on by $H$ results in the state $HX\ket{\Psi}=HXHH\ket{\Psi}=ZH\ket{\Psi}$. This can be verified by simple matrix multiplication. In other words, an $X$ error commuted through an $H$ gate transforms into a $Z$ error. Similarly,
\begin{eqnarray}
& Z \stackrel{H}{\longrightarrow} X, & \\
& I\otimes X \stackrel{C_X}{\longrightarrow} I\otimes X, & \label{ix} \\
& X\otimes I \stackrel{C_X}{\longrightarrow} X\otimes X, & \label{xi} \\
& I\otimes Z \stackrel{C_X}{\longrightarrow} Z\otimes Z, & \\
& Z\otimes I \stackrel{C_X}{\longrightarrow} Z\otimes I. &
\end{eqnarray}
For brevity, where clear from context, outer products such as $X\otimes I$ are frequently written as $XI$. We shall use the above rules for propagating errors extensively.

\begin{figure}
\begin{center}
\resizebox{55mm}{!}{\includegraphics{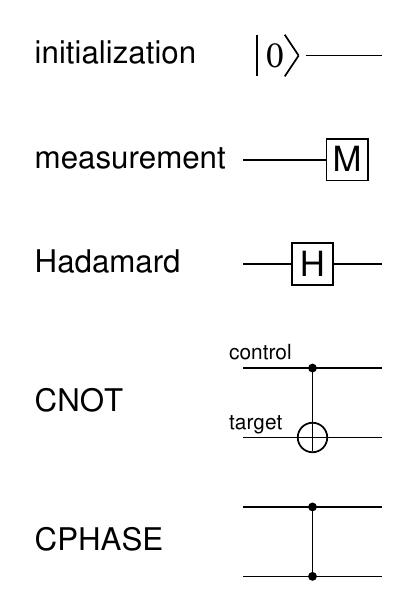}}
\end{center}
\caption{Quantum circuit symbols for initialization, measurement, Hadamard, CNOT and CPHASE.}\label{definitions}
\end{figure}

\begin{figure}
\begin{center}
\resizebox{90mm}{!}{\includegraphics{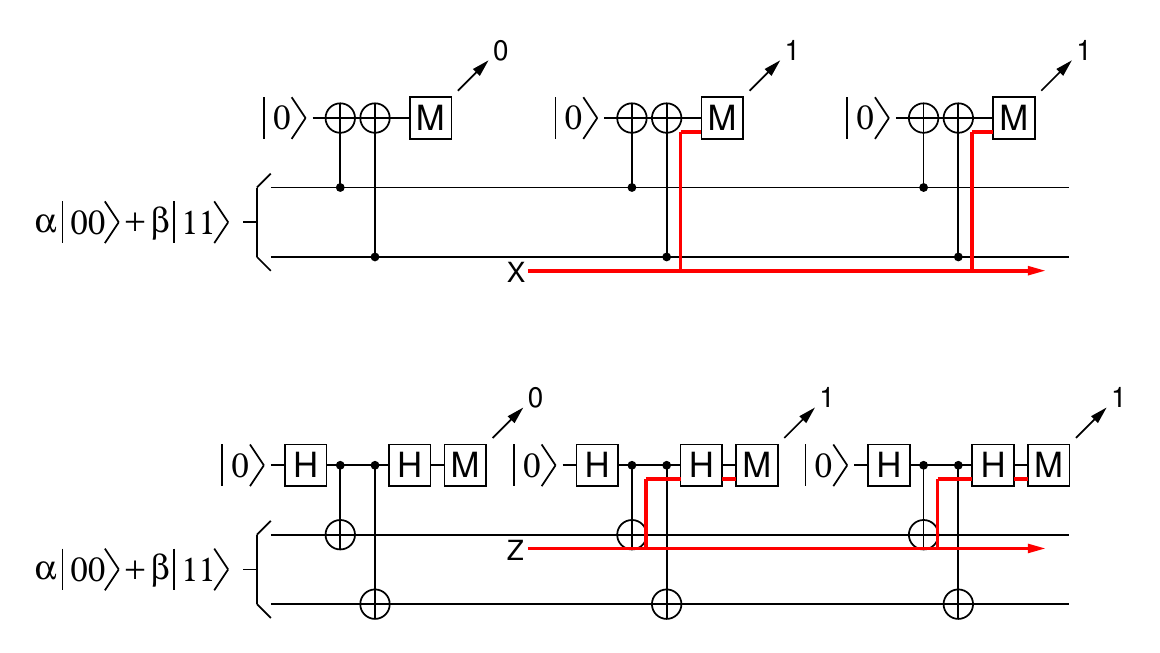}}
\end{center}
\caption{Examples of error detection circuits. Errors lead to a permanent change in the periodic circuit measurement value. A detection event is associated with any sequential pair of measurements that differ in value.}\label{detection}
\end{figure}

Quantum circuits provide a convenient notation for expressing complex sequences of quantum gates. Fig.~\ref{definitions} defines common circuit symbols. Two simple periodic quantum circuits are shown in Fig.~\ref{detection}. When the measurement value of these periodic circuits changes value, the local presence of one or more errors is indicated. These measurement value changes are called detection events. Two similar (sometimes identical) but conceptually distinct circuits are always required to detect all quantum errors requiring correction. We arbitrarily label one of these primal and the other dual. Primal (dual) detection circuits lead to primal (dual) detection events. Errors that lead to primal (dual) detection events are called primal (dual) errors.

\section{Topological quantum error correction}
\label{TQEC}

\begin{figure}
\begin{center}
\resizebox{85mm}{!}{\includegraphics{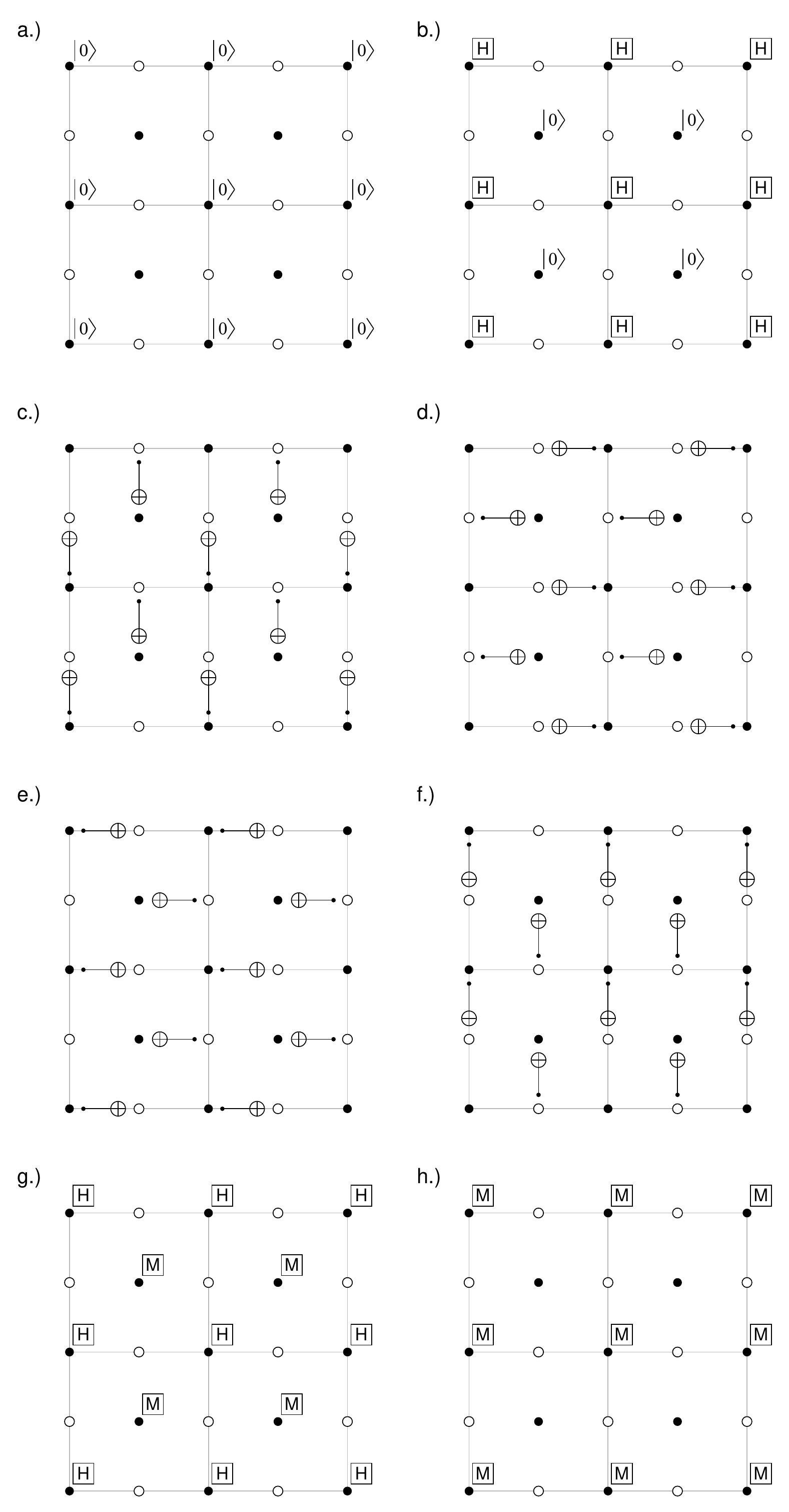}}
\end{center}
\caption{An eight layer sequence of quantum gates comprising a single round of surface code quantum error detection. White circles represent data qubits, black circles represent syndrome (error detection) qubits. Grey edges are guides for the eye with no physical significance. Definitions of all quantum gates are given in Fig.~\ref{definitions}.}\label{sequence}
\end{figure}

\begin{figure}
\begin{center}
\resizebox{85mm}{!}{\includegraphics{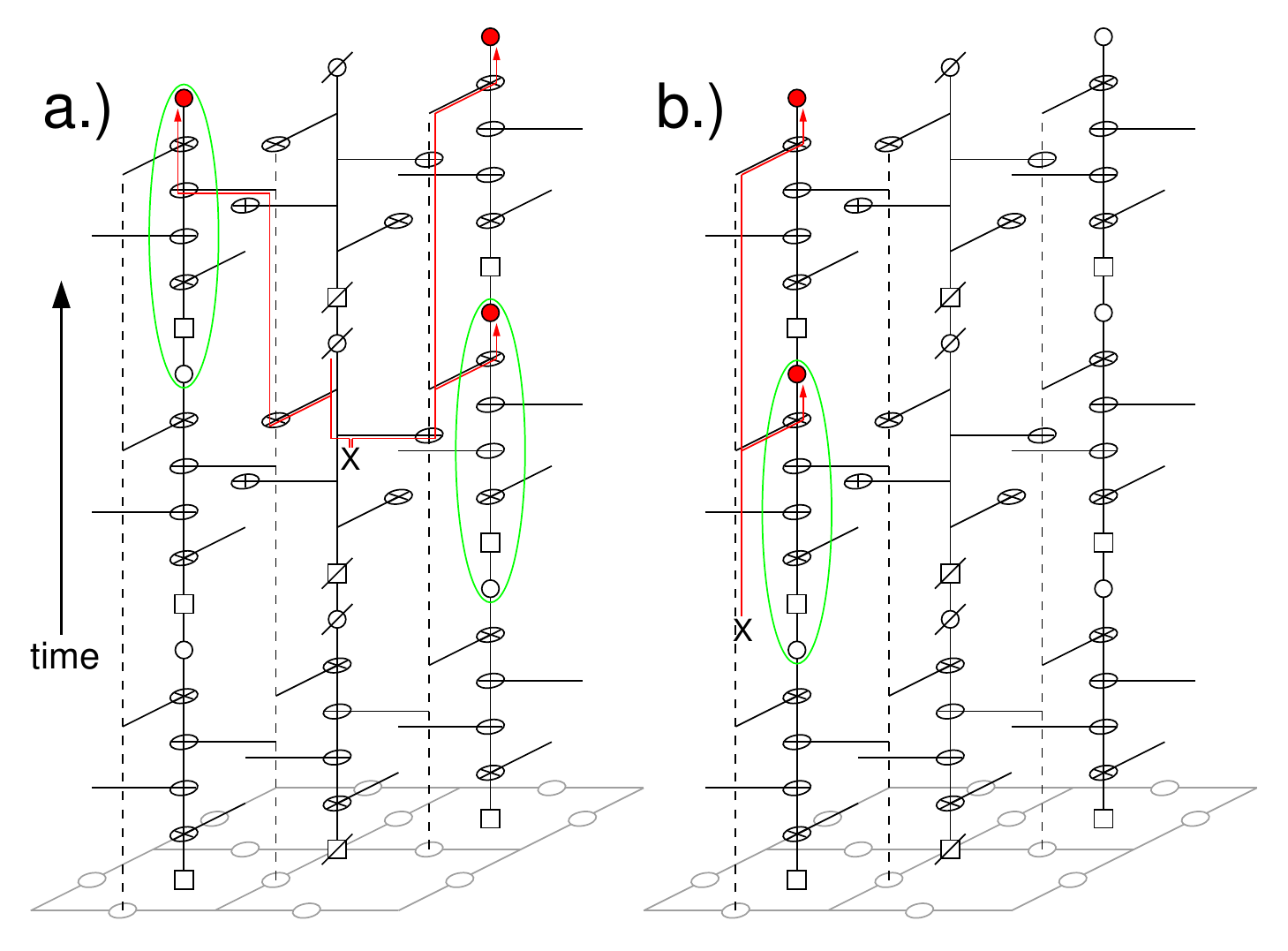}}
\end{center}
\caption{2-D surface code (grey). Time runs vertically. Squares represent initialization to $\ket{0}$, circles represent $Z$ basis measurement. Slashed squares represent initialization to $\ket{+}$, slashed circles represent $X$ basis measurement. a.) A single error leading to a pair of detection events (green ellipses encircle each pair of measurements with differing value associated with each detection event). Red lines show the paths of error propagation, using Eq.~\ref{ix}. b.) An error leading to a single detection event due to proximity to a boundary of the lattice.}\label{syn_err}
\end{figure}

For the purposes of this paper, topological quantum error correction (TQEC) is defined to be a collection of quantum circuitry on an arbitrary dimensional, nearest neighbor coupled lattice of qubits with the property that a single error leads to a pair of primal and/or dual detection events unless the error is near a boundary of the lattice. Near boundaries, a single error can lead to just a single detection event. A 2-D planar circuit with these properties is shown in Fig.~\ref{sequence}. This circuit is associated with the surface code \cite{Brav98,Denn02,Raus07,Raus07d,Fowl08,Fowl12f}. Examples of errors resulting in a pair of detection events and a single detection event are shown in Fig.~\ref{syn_err}. Note that error correction codes have traditionally been defined in terms of stabilizers \cite{Gott97}, however our definition in terms of quantum circuits is both more general and how Autotune works internally.

\begin{figure}
\begin{center}
\resizebox{75mm}{!}{\includegraphics{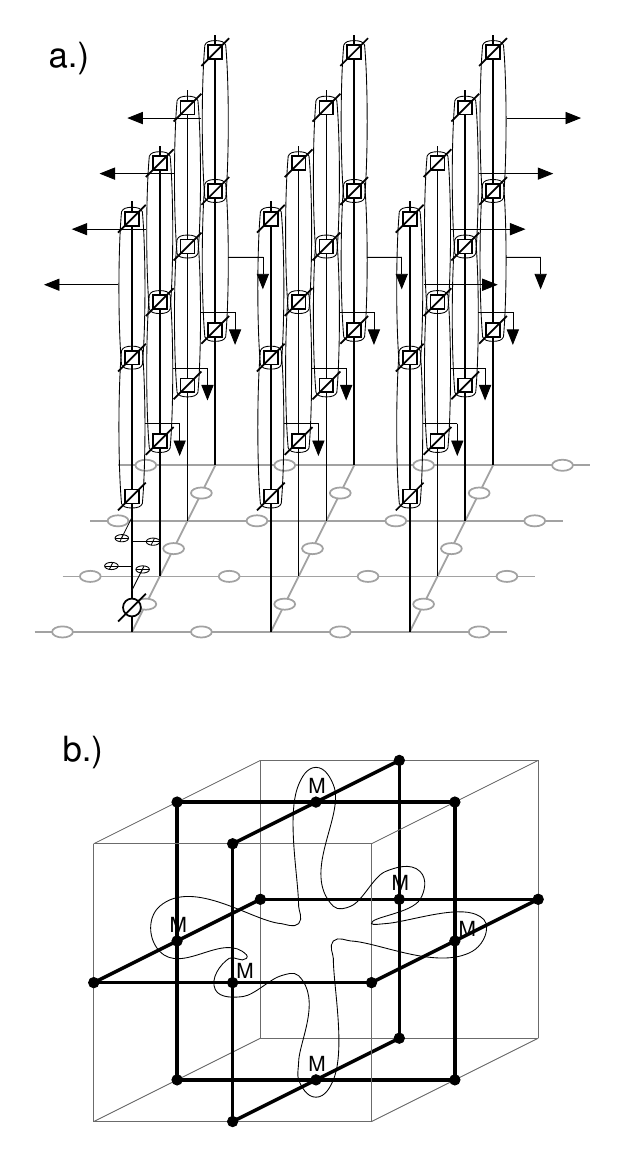}}
\end{center}
\caption{a.) 2-D surface code (grey). Time runs vertically. In the front left corner, the transversely invariant initialization, 4 CNOT, then measurement pattern is shown. For clarity, only the measurement gate of this pattern is shown elsewhere in space-time. Long ovals encircle pairs of measurements and represent sets. Arrows indicate associations with boundaries. The bottom layer of sets is associated with the initial time boundary. The second layer left and right rows of sets are associated with the left and right spatial boundaries respectively. b.) A single 3-D topological cluster state cell. Sets contain six measurements away from boundaries of the lattice. Dots represent qubits initialized to $\ket{+}$, heavy black lines connecting dots represent CPHASE gates. The order of application of CPHASE gates is irrelevant as they commute.} \label{sets}
\end{figure}

The notion of a detection event as defined above can be significantly generalized. Defining measurement to report the results +1 and -1 for $\ket{0}$ and $\ket{1}$ respectively, a detection event can be associated with a set of measurements with product -1. Examples of appropriate sets are shown in Fig.~\ref{sets}a. This figure also contains arrows indicating nearby temporal and spatial boundaries. This information, sets and boundaries, is required by Autotune to correctly construct detection events and associate single detection events with the correct boundary.

Armed with these definitions, we can also consider more complex TQEC schemes, such as 3-D topological cluster states \cite{Raus06,Raus07d,Fowl09}. A cluster state \cite{Raus01,Raus03} can be prepared in two stages by first initializing a number of qubits to $\ket{+}=(\ket{0}+\ket{1})/\sqrt{2}$. Physically, this typically corresponds to initializing the qubits to $\ket{0}$ and applying Hadamard. Next controlled-phase (CPHASE or $C_Z$) gates are applied to pairs of qubits, where
\begin{eqnarray}
C_Z & = & \left( \begin{array}{cccc} 1 & 0 & 0 & 0 \\ 0 & 1 & 0 & 0 \\ 0 & 0 & 1 & 0 \\ 0 & 0 & 0 & -1 \end{array} \right).
\end{eqnarray}
A topological cluster state has the form shown in Fig.~\ref{sets}b. This basic structure is tiled in 3-D.

When all topological cluster state qubits are measured in the $X$ basis, typically achieved by applying Hadamard and then measuring, errors can be detected by multiplying the measurement results in sets of the form shown in Fig.~\ref{sets}b. As above, a -1 product indicates a detection event. This is discussed in more detail in \cite{Fowl09}. Exactly why this is true is not important for the purposes of this paper. The only feature of TQEC required to read this paper is that detection events are associated with particular locations in space-time and that single errors lead to detection events that are local to one another.

\begin{figure}
\begin{center}
\resizebox{80mm}{!}{\includegraphics{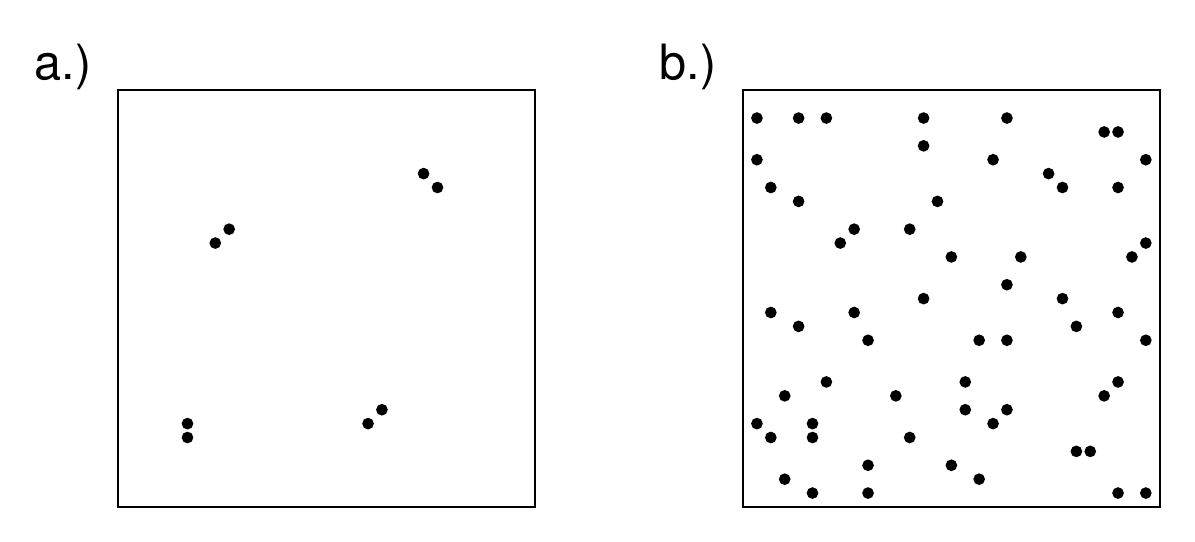}}
\end{center}
\caption{a.) Typical low error rate pattern of detection events. Such patterns are easy to correct. b.) Typical high error rate pattern of detection event with significant probability of unsuccessful correction.}\label{patterns}
\end{figure}

If two errors would lead to a detection event at the same location, their contributions cancel and no detection event is observed at the shared location. Instead, an error chain is formed, with detection events only at the endpoints of the chain. Assuming some characteristic error rate $p$ and independent errors, long error chains are exponentially unlikely. Assuming a low error rate, a typical pattern of detection events consists of well isolated pairs as shown in Fig.~\ref{patterns}a. The higher the error rate, the more difficult it is to guess the location of the errors leading to the detection events. A high error rate pattern of detection events is shown in Fig.~\ref{patterns}b.

A number of different algorithms exist taking detection events in topological codes and determining locations to apply corrections \cite{Wang09b,Ducl09,Ducl10,Land11,Bomb12,Woot12,Sarv12}. Edmonds' minimum weight perfect matching algorithm \cite{Edmo65a,Edmo65b,Fowl11b,Fowl12c} is the fastest known algorithm capable of handling practical circuits of the form described in this Section. Minimum weight perfect matching is conceptually simple --- after connecting detection events in pairs or individually to nearby boundaries such that the total length of connecting paths is minimal, corrections are applied along the connecting paths. The errors leading to the detection events in Fig.~\ref{patterns}a would be corrected with high probability, those in Fig.~\ref{patterns}b may not be successfully corrected. The error rate at which adding additional qubits fail to improve the probability of successful correction is called the threshold error rate $p_{\rm th}$. In minimal gate count fault-tolerant implementations of the surface code, simulations indicate $p_{\rm th}\sim1\%$ \cite{Wang11,Fowl11b}, consistent with the proven lower bound to the threshold error rate of $7.4\times 10^{-4}$ \cite{Fowl12e}.

Autotune uses our own implementation of minimum weight perfect matching \cite{Fowl11b,Fowl12c}. The details of this implementation lie outside the scope of this discussion as Autotune itself treats matching as a black box process. The primary question we wish to address is how to define the distance between two detection events. If we assign coordinates to detection events such that neighboring events $v_1$, $v_2$ of the same type (primal or dual) differ by one unit in one coordinate, the default metric is the Manhattan metric
\begin{eqnarray}
D(v_1,v_2) & = & |i_1-i_2|+|j_1-j_2|+|k_1-k_2|.
\end{eqnarray}
This metric was used in early TQEC works \cite{Raus07,Fowl08,Wang09}.

\section{Tracking and visualizing errors}
\label{Autotune}

\begin{figure}
\resizebox{60mm}{!}{\includegraphics{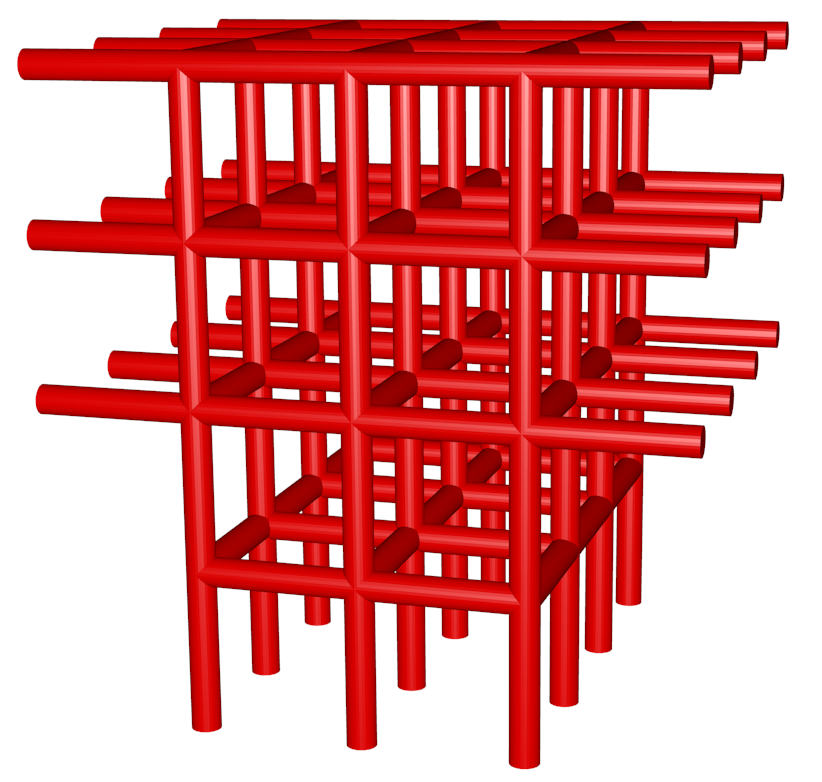}}
\caption{Manhattan primal lattice of a distance 4 surface code with depolarizing noise. Time runs vertically. Cylinder endpoints represent points in space-time where detection events can occur. Each cylinder has equal diameter, representing an assumed equal probability of detection events at the endpoints of each cylinder. Vertical cylinders leading to nowhere connect to the initial time boundary. Horizontal cylinders leading to nowhere connect to spatial boundaries of the qubit array.}
\label{Manhattan_sc}
\end{figure}

The Manhattan metric takes no details of the underlying gate sequence or error models of each gate into account. If we imagine every potential location of a primal or dual detection event in space-time, namely the location of each primal or dual set, the Manhattan metric can be visualized as a cubic lattice. Fig.~\ref{Manhattan_sc} shows the Manhattan lattice associated with the primal sets shown in Fig.~\ref{sets}a. Each cylinder represents a weight 1 path. A similar lattice exists for the dual sets. The distance $D$ between any given pair of sets is defined to be the shortest (lowest weight) connecting path through the corresponding lattice. Cylinders apparently leading to nowhere actually lead to spatial or temporal boundaries. Given a lattice and randomly generated detection events, minimum weight perfect matching can be used to match detection events to one another or nearby boundaries such that the total weight of all connecting paths through the lattice is minimal.

Manhattan lattices are trivial to construct for the surface code and topological cluster states. However, as we shall see in detail in Section~\ref{versus}, Manhattan lattices lead to very suboptimal performance. The reason for this can be deduced from Fig.~\ref{syn_err}a, in which a single error can be seen to lead to a pair of detection events separated by two units of space and one unit of time. In other words, when using a Manhattan lattice, this single error is treated the same way as some three error chains. In \cite{Fowl10}, we showed that this shortcoming leads to a given size surface code only being able to guaranty correction of half the number of errors it is theoretically capable of always correcting. This shortcoming was removed by laboriously hand analyzing the surface code circuits and error models and including diagonal weight 1 links (additional diagonal cylinders with the same diameter) in the lattice wherever necessary.

Weight 1 links do not take into account the relative probability of different pairs of detection events. Generally speaking, diagonally separated pairs of detection events are less probable than pairs separated only along one axis of the cubic lattice. In \cite{Wang11}, with even more laborious hand analysis, all distinct types of error were propagated through the surface code and polynomial expressions were obtained for the probability of each link as a function of a characteristic physical gate error rate $p$. This analysis was performed only for the body of the lattice, and not attempted near the boundaries. Links to the spatial boundaries were simply assigned the same probability as a horizontal link of appropriate direction from the body of the lattice. It was deemed too complex and laborious to perform this analysis for the different types of boundaries and corners of the lattice.

\begin{figure}
\resizebox{60mm}{!}{\includegraphics{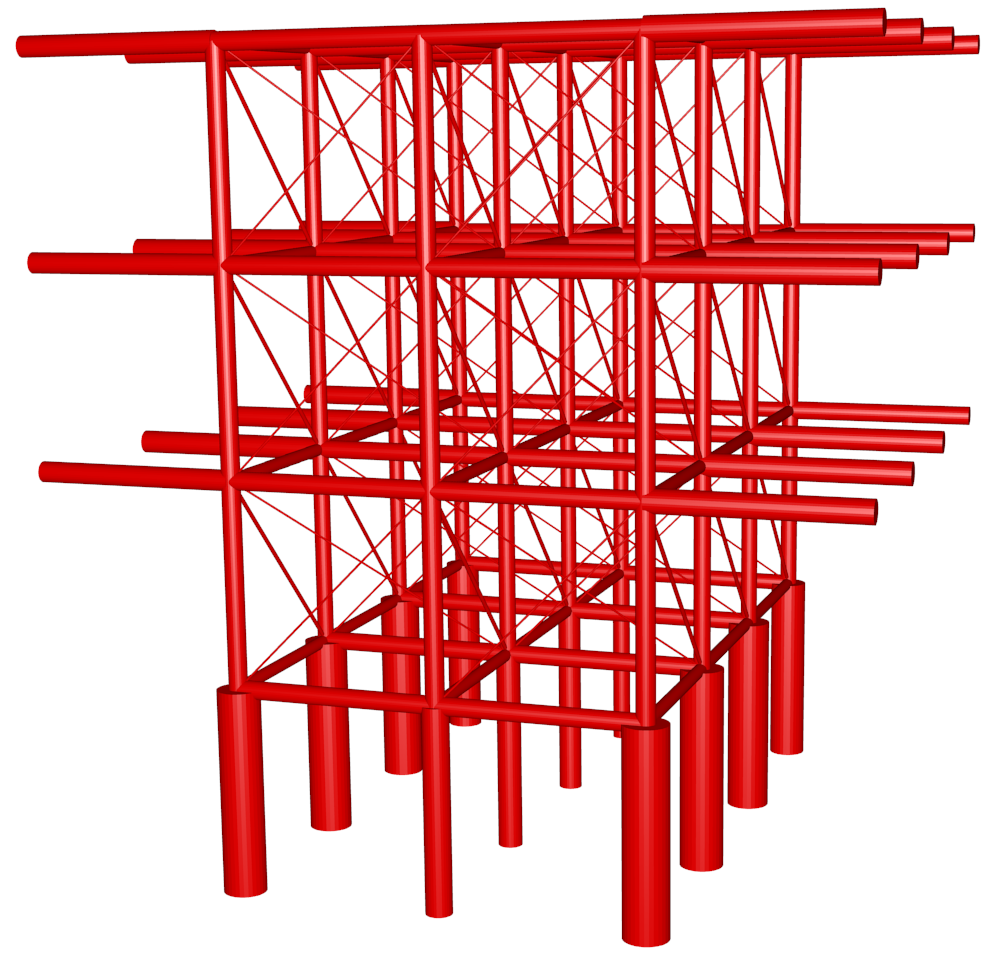}}
\caption{Autotune generated primal lattice of a distance 4 surface code with depolarizing noise. Note that in contrast to Fig.~\ref{Manhattan_sc}, cylinders do not have equal diameter, accurately representing the diverse range of probabilities of various pairs of detection events. Note also the many additional diagonal cylinders, which are associated with errors that propagate to space-time location separated by more than one unit of space and/or time (see Fig.~\ref{syn_err}).}
\label{Autotune_sc}
\end{figure}

When given an arbitrary TQEC circuit and arbitrary stochastic error models for each gate, Autotune performs a full analysis of the propagation of all errors through all parts of the circuit to determine the probability of all pairs of local detection events. As this is not conceptually complex, we shall not describe the details of how is this is achieved here. A description of the Autotune algorithm can be found in Appendix~\ref{alg}. The analysis typically takes less than one second. A lattice is then constructed with link weights equal to $-ln(p_{\rm link})$, where $p_{\rm link}$ is the total probability of all errors leading to a given link, discussed in more detail in Appendix~\ref{Nests}. This ensures that low probability links have larger positive weights and are therefore used less often. It also ensures that the sum of weights along a path through the lattice is related to the product of probabilities along that path. Fig.~\ref{Autotune_sc} shows the Autotune generated lattice corresponding to Fig.~\ref{sets}a. Cylinders now have a diameter proportional to the probability of that link, which we find more useful than the weight for visualization purposes. Note the varying diameters of cylinders to the boundary.

\begin{figure}
\resizebox{60mm}{!}{\includegraphics{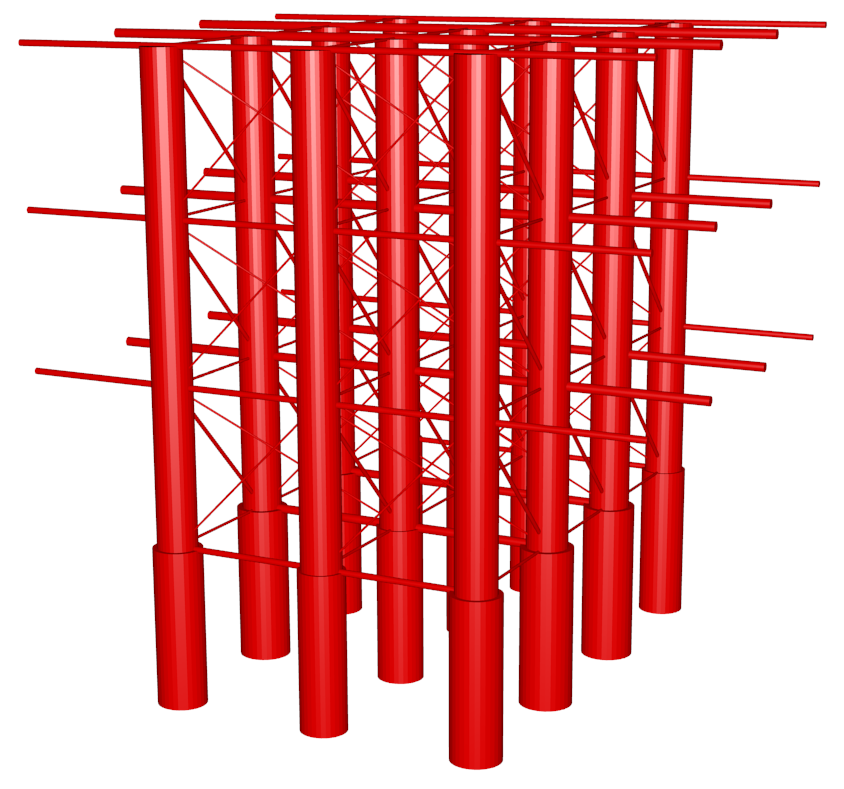}}
\caption{Autotune generated primal lattice of a distance 4 surface code with measurement error rate $10p$, identity error rate $0.1p$ and a CNOT with total probability of error $p$ but with any error containing $Y$ or $Z$ 100 times more likely than an error containing only $I$ and $X$. Note that the increased measurement error probability has led to fat vertical cylinders, representing the increased probability that consecutive measurements will differ leading to vertically aligned detection events. Such vertically aligned detection events will be preferentially matched together instead of to nearby boundaries.}
\label{Asymmetric_sc}
\end{figure}

Fig.~\ref{Autotune_sc} was created with a standard depolarizing error model. Initialization and measurement produce or report the wrong state with probability $p$, identity and Hadamard gates introduce an $X$, $Y=XZ$ or $Z$ error each with probability $p/3$, and CNOT introduces one of the 15 nontrivial outer products of $I$, $X$, $Y$ and $Z$ each with probability $p/15$. Fig.~\ref{Asymmetric_sc} shows a lattice generated with measurement error rate $10p$, identity error rate $0.1p$ and a CNOT with total probability of error $p$ but with any error containing $Y$ or $Z$ 100 times more likely than an error containing only $I$ and $X$. Note the much thicker vertical cylinders due to the high measurement error rate. Autotune is designed to handle absolutely any set of stochastic error models. Note that after the first few layers the structure of the lattice repeats. Exactly the same pattern of links and probabilities will be generated since the TQEC circuit is repetitive. This means that we do not need to endlessly perform an analysis of the propagation of all errors, rather we perform the analysis until it becomes repetitive, then simply continue to generate the repetitive structure without the analysis.

\section{Unoptimized versus Autotuned TQEC}
\label{versus}

When we simulate the surface code, we study the situation of a lattice of finite spatial extent and potentially infinite temporal extent. In practice, the lattice is generated dynamically as gates are simulated and random errors generated. We are interested in the logical error rate per round of error detection. A logical error is a chain of errors after correction that connects distinct boundaries. These errors can be detected in simulations as changes in the logical state.

The distance $d$ of a code is the minimum number of operations required to change the logical state. For each $(d,p)$ pair, we typically let our simulation run until 10,000 logical state changes are observed. Fewer state changes are occasionally permitted at very low $p$ and high $d$ as the runtime of the simulations increases rapidly due to the very low logical error rate and consequent large number of required rounds of simulated error detection. We have recently developed fast analytic methods to calculate the low $p$ logical error rate for arbitrary even $d$ \cite{Fowl12g}.

\begin{figure}
\begin{center}
\resizebox{85mm}{!}{\includegraphics[viewport=60 60 545 430, clip=true]{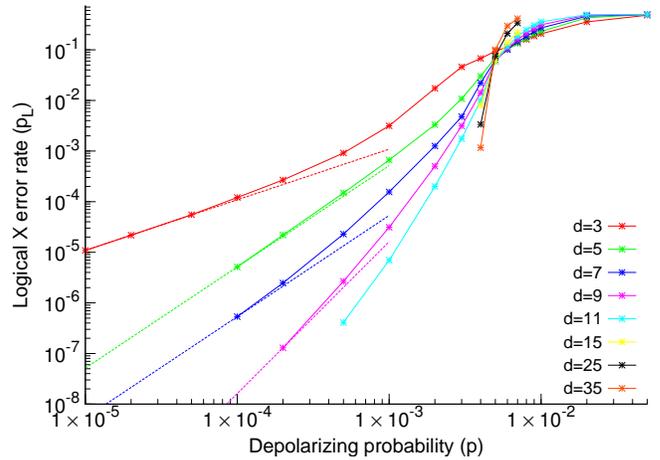}}
\end{center}
\caption{Probability of logical $X$ error per round of error correction for various code distances $d$ and physical error rates $p$ when using a Manhattan lattice. The asymptotic curves (dashed lines) are linear, quadratic, quadratic, cubic for distances $d=3,5,7,9$ respectively.}\label{logx_sc}
\end{figure}

\begin{figure}
\begin{center}
\resizebox{85mm}{!}{\includegraphics[viewport=60 60 545 430, clip=true]{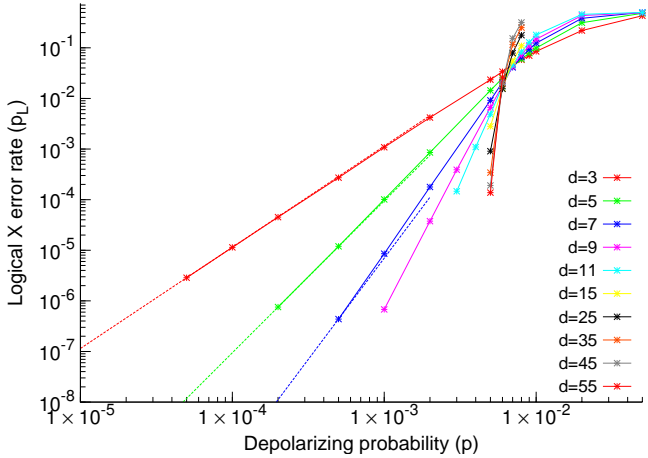}}
\end{center}
\caption{Probability of logical $X$ error per round of error correction for various code distances $d$ and physical error rates $p$ when using an Autotune generated lattice. The asymptotic curves (dashed lines) are quadratic, cubic, quartic for distances $d=3,5,7$ respectively.}\label{logx_at_sc}
\end{figure}

The logical $X$ error rate per round of error correction for a range of values of $d$ and $p$ when using Manhattan lattices is shown in Fig.~\ref{logx_sc}. This should be contrasted with the performance when using Autotune generated lattices, shown in Fig.~\ref{logx_at_sc}.

The parallel asymptotic curves for distances 5 and 7 in Fig.~\ref{logx_sc} are correct. In Fig.~\ref{syn_err}a, an example of a single error leading to detection events are separated by two units of space and one unit of time is shown. When using the Manhattan lattice, three corrections must be inserted to pair such detection events. Said another way, when using the Manhattan lattice, this single error is indistinguishable from a three error process. A logical error occurs when detection events are incorrectly matched forming a logical operator of errors and corrections. Logical operators are associated with paths through the lattice connecting opposing boundaries. The shortest paths make use of the same number of links as the code distance. There exist logical operators in a $d=5$ code that do not follow a shortest path and are associated with of a total of 9 links and incorporate two single-triple errors. Given the detection events associated with these single-triple errors, matching will choose to match incorrectly (inserting 3 corrections) rather than correctly (which would require 6 corrections), forming a logical error. This is why the asymptotic curve for $d=5$ is only quadratic. Similarly, a $d=7$ code contains logical operators consisting of 11 links incorporating two single-triple errors and matching will choose to match incorrectly (inserting 5 corrections) rather than correctly (6 corrections). Fewer combinations of two single-triple errors lead to failure in a $d=7$ code, explaining the lower logical error rate. Autotune lattices do not suffer the shortcomings described in this paragraph as they include all necessary diagonal links and therefore guarantee correction of $\lfloor (d-1)/2 \rfloor$ errors.

If we focus on physical error rates of $p=10^{-3}$, a reasonable medium-term goal for scalable two-qubit interactions, the distance $d=5$ Autotuned logical error rate is nearly a factor of 10 lower than the Manhattan logical error rate. At $d=7$, the improvement is over a factor of 40. The ratio of outperformance continues to grow with $d$. The ratio of outperformance also grows rapidly as $p$ is decreased. At $p=10^{-5}$, the outperformance is already a factor of a hundred at $d=3$, and over five orders of magnitude at $d=7$. This extreme performance difference makes Autotune an essential tool for analyzing the surface code.

\begin{figure}
\resizebox{60mm}{!}{\includegraphics{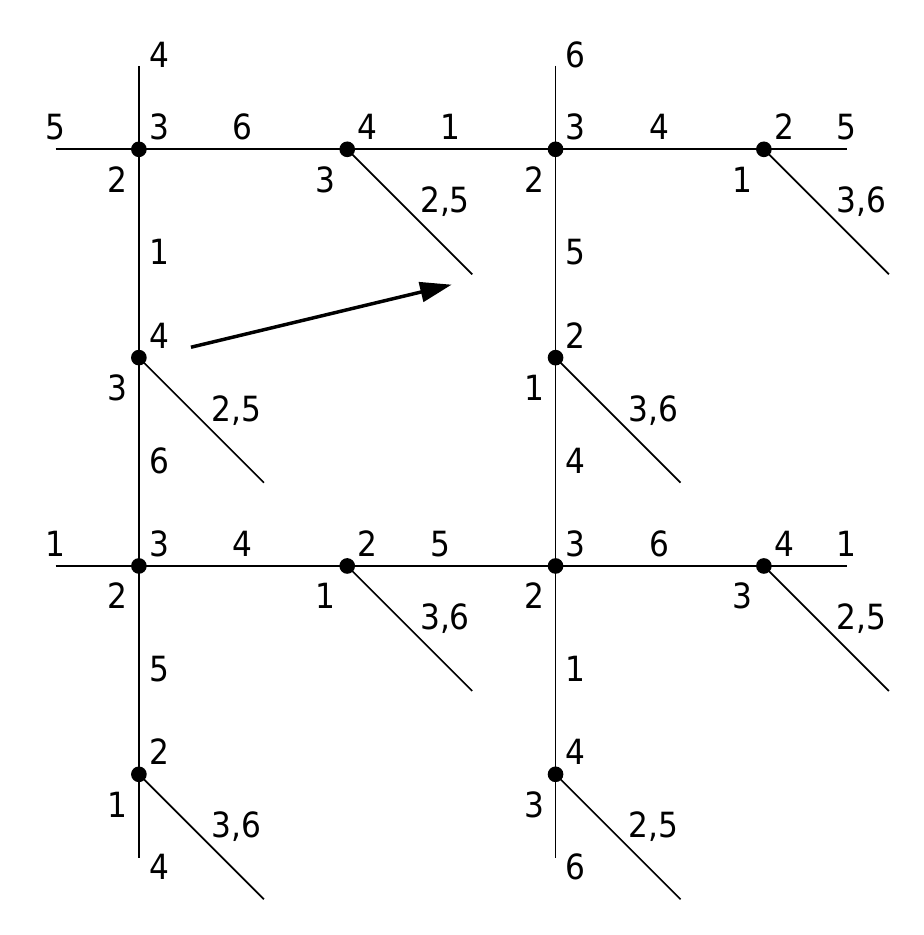}}
\caption{A unit cell of the first layer of the interaction pattern of a double layer of qubits generating a 3-D cluster state. Small black dots correspond to qubits. The qubits in the second layer have been omitted for clarity. The numbers at the top right and bottom left of each qubit indicate the time steps in which it is initialized and measured, respectively. Each line corresponds to a $C_Z$ gate and is labeled with the time step(s) in which this gate is applied. The second layer comprises an identical ordering of qubits and gates, but shifted in time by three steps (such that a gate labeled 2 is executed in time step 5 and vice versa, etc.) and shifted north, east and up as indicated by the arrow. Upward $C_Z$ gates in the first layer become downward after shifting.}\label{cs_1layer}
\end{figure}

\begin{figure}
\resizebox{60mm}{!}{\includegraphics{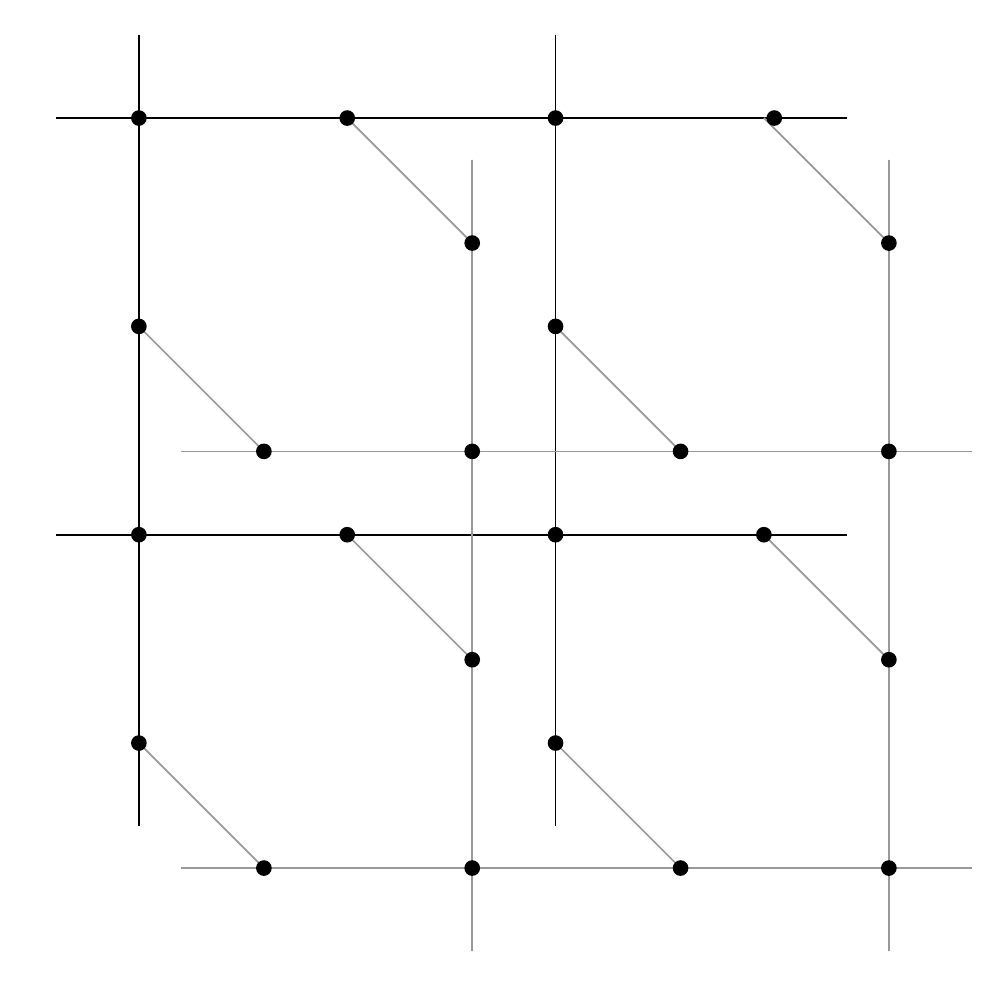}}
\caption{A single-layer implementation of the two layers used to progressively generating a 3-D topological cluster state. Note that while some interactions cross, they are still technically nearest neighbor.}\label{cs_2layers}
\end{figure}

Turning our attention to topological cluster states, the first step is to describe a way to progressively build the cluster state using only a 2-D lattice of qubits and nearest neighbor interactions. We start with a double layer of qubits. The required sequence of initialization, measurement, and $C_Z$ gates is somewhat complex, however an attempt to convey this clearly can be found in Fig.~\ref{cs_1layer}. An appropriate single layer of qubits and interactions capable of implementing the same sequence is shown in Fig.~\ref{cs_2layers}.

\begin{figure}
\resizebox{60mm}{!}{\includegraphics{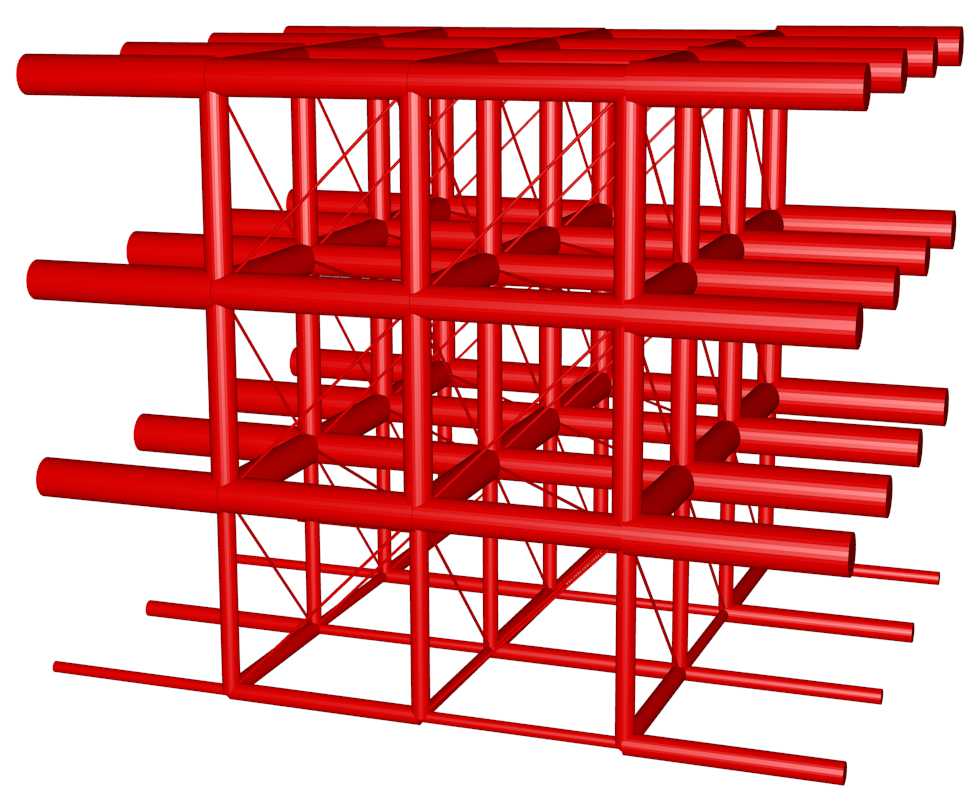}}
\caption{Autotune generated primal lattice of a distance 4 topological cluster state with depolarizing noise.}
\label{cluster}
\end{figure}

\begin{figure}
\begin{center}
\resizebox{85mm}{!}{\includegraphics[viewport=60 60 545 430, clip=true]{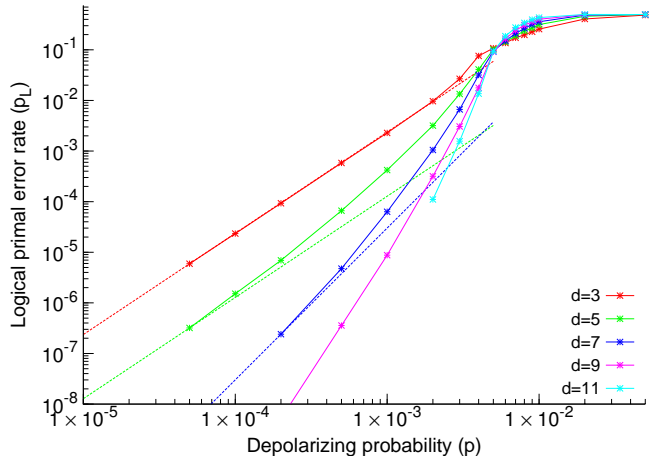}}
\end{center}
\caption{Probability of logical primal error per round of error correction for various topological cluster state code distances $d$ and physical error rates $p$ when using a Manhattan lattice. The asymptotic curves (dashed lines) are quadratic, quadratic, cubic for distances $d=3,5,7$ respectively.}\label{log_pr}
\end{figure}

\begin{figure}
\begin{center}
\resizebox{85mm}{!}{\includegraphics[viewport=60 60 545 430, clip=true]{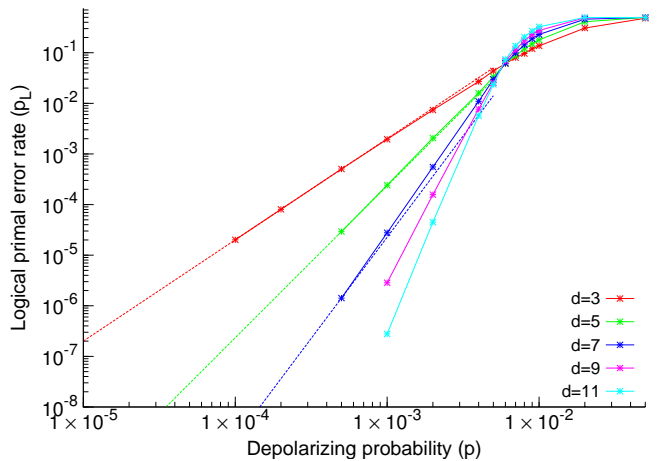}}
\end{center}
\caption{Probability of logical primal error per round of error correction for various topological cluster state code distances $d$ and physical error rates $p$ when using an Autotune generated lattice. The asymptotic curves (dashed lines) are quadratic, cubic, quartic for distances $d=3,5,7$ respectively.}\label{log_pr_at}
\end{figure}

We assume the same error model as that used for the surface code, including the explicit use of Hadamards gates and initialization and measurement in the $Z$ basis only. The primal nest resulting from the described gate sequence and error model for a distance 4 topological cluster state is shown in Fig.~\ref{cluster}. Figs.~\ref{log_pr}--\ref{log_pr_at} showed the Manhattan lattice and Autotuned lattice error correction performance, with the distance 5 logical error rate in particular showcasing the advantage of using Autotune. Note the parallel asymptotic curves for $d=3$ and 5 in the Manhattan case, caused by single errors leading to detection events separated by two links.

\section{Conclusion}
\label{Conclusion}

We have described a tool Autotune that is capable of handling in a natural manner both fully fault-tolerant surface codes and 3-D topological cluster states, with the full code distance achieved in both cases. This generality is achieved through the definition of sets of measurements which are used to detect error chain endpoints. Arbitrary syndrome measurement circuits are supported along with arbitrary stochastic error models for each gate. The details of the measurement circuits and gate error models are analyzed before simulation begins to ensure high error correction performance. The algorithms upon which Autotune is based are highly efficient, with per round runtime comparable to that reported in \cite{Fowl12c}.

In future work, we plan to take correlations between errors into account, improve the efficiency of Autotune's handling of qubit loss, parallelize the core matching engine, and develop the capability to analyze complex logical circuits consisting of many braided defects, including the precise simulation of the complete quantum state.

\section{Acknowledgements}

This research was conducted by the Australian Research Council Centre of Excellence for Quantum Computation and Communication Technology (project number CE110001027), with support from the US National Security Agency and the US Army Research Office under contract number W911NF-08-1-0527. Supported by the Intelligence Advanced Research Projects Activity (IARPA) via Department of Interior National Business Center contract number D11PC20166.  The U.S. Government is authorized to reproduce and distribute reprints for Governmental purposes notwithstanding any copyright annotation thereon.  Disclaimer: The views and conclusions contained herein are those of the authors and should not be interpreted as necessarily representing the official policies or endorsements, either expressed or implied, of IARPA, DoI/NBC, or the U.S. Government.

\bibliography{../References}

\appendix

\section{Two-qubit experiments demonstrating TQEC}
\label{2q_expt}

This Appendix describes generic experiments requiring as few as two qubits that are sufficient to prove that a physical system is capable of implementing TQEC. A large 2-D array of qubits is not required provided the experiment is done in a scalable manner. It must be clear where an arbitrarily large number of additional qubits will go, where all of the control lines will go, where all of the control electronics will go, scalable cooling if required, and scalable construction methods for the complete large-scale hypothetical quantum computer. No issues can be ignored. This is best achieved with a modular design, such as that described in \cite{Devi08}.

The simplest possible modular structure capable of being assembled into a large 2-D array of qubits contains just a single qubit with the ability to connect to four neighboring modules. Two such modules connected together are sufficient to demonstrate initialization, measurement, identity, single- and two-qubit gates. If the error rates of all of these quantum gates are below approximately 1\%, the experiment will have demonstrated that nature permits large-scale quantum computation. We believe that the importance of such an experiment cannot be overstated.

A number of trade-offs are possible. For example, measurement error rates of approximately 10\% are acceptable provided two-qubit interactions, CNOT or CZ, are possible with error rate approximately 0.1\%. Precise trade-offs can be determined for specific hardware using Autotune.

Each module need not be a physically separate device. A pair of solid-state modules may consist of a pair of coupled qubits together with coupling hardware for the six additional nearest neighbor qubits, which may or may not themselves actually be present. At least the coupling hardware needs to be present to ensure that control and decoherence challenges associated with scaling have indeed been overcome. It is possible that additional qubits and coupling hardware may need to be present if the mechanism achieving qubit interactions significantly couples more than just nearest neighbor qubits.

To take a specific example, consider a \emph{scalable} ion trap quantum computer based on a regular array of interaction regions connected by a square grid of transport paths. Provided the design is truly scalable, it would be sufficient to build just two unit cells of the scalable array containing in total perhaps just two interaction regions and two transport junctions. Four ions would be required, two data ions and two sympathetic cooling ions. One would then need to demonstrate high fidelity (>0.99) initialization, measurement and single-qubit rotations. Most challengingly, one would need to demonstrate an interaction process consisting of transport of one ion to the other interaction region, sympathetic cooling if necessary, combination, high fidelity interaction, separation, transport, more sympathetic cooling if necessary, and prove that the entire process had fidelity >0.99. One would also need to show that an identity gate of duration equal to the interaction process could be achieved with fidelity >0.99. Finally, one would need to show that ion loss and leakage to non-computational states were low probability events, preferably less than 1\% per interaction process or other gate. An experiment of this form would put a price tag on a qubit and enable one to calculate, using Autotune, the precise size, cost and performance of a large-scale quantum computer based on many of the experimentally demonstrated unit cells. This would reduce the problem of building a quantum computer to one of price-performance, a problem sure to attract significant industry interest.

We would be delighted to have arbitrarily detailed discussions with anyone wishing to design a ``nature permits quantum computation'' experiment in any type of hardware.

\section{Autotune algorithm}
\label{alg}

The purpose of Autotune is to enable the precise analysis of real hardware running topological quantum error correction (TQEC). This is achieved by taking user-defined error models for every quantum gate and user-defined quantum circuits implementing TQEC and working out exactly where every possible error on every gate would be detected in space-time and which errors lead to the same detection events. When simulating the operation of the hardware stochastically, the detailed error propagation information produced by Autotune can be used to reliably guess which errors led to the observed detection events. This results in extremely effective correction of physical errors. In this Appendix, we describe how Autotune represents error models, how these errors are propagated through quantum circuits, how errors are detected (sets of measurements), how the detailed error information is visualized (nests of balls and sticks), and how the detailed error information can be reduced to simpler structures that can be generated efficiently during stochastic simulation (lattices of dots and lines).

\subsection{Error models and tracking}
\label{Error models}

Autotune is capable of handling any error model with outcomes that can be described by a single integer $e$ per qubit. For example, $I=0$, $X=1$, $Z=2$, $Y=3$, leaked to a non-computational state = 4, qubit lost = 5. Only single-qubit and two-qubit gate error models are currently supported, however this could easily be extended. The user can specify how $e$ is transformed by each gate, for example controlling whether a CNOT between a leaked qubit and a non-leaked qubit results in two leaked qubits, or no effect on the non-leaked qubit, or any other effect describable by single integers.

An example of a Pauli channel CNOT error model file is shown below.
\begin{eqnarray}
&&2\nonumber\\
&&1.0\nonumber\\
&&6\nonumber\\
&&81~~0~~3\nonumber\\
&&14~~1~~0\nonumber\\
&&13~~2~~1\nonumber\\
&&28~~2~~2\nonumber\\
&&78~~3~~0\nonumber\\
&&30~~3~~2\nonumber\\
&&1
\end{eqnarray}
The first line states the number of qubits $n_q$ the gate is applied to. The second line states the value $x$ the relative strengths $s_i$ of the various errors should be normalized to sum to. This makes it easy to handle different gates with different overall probabilities of error, e.g.~$x_{\rm 1q}=x_{\rm 2q}/10$. The third line states the number of different errors $n_e$ in the model. The next $n_e$ lines contain $n_q+1$ integers specifying the relative strength $s_i$ of that error and the value of $e$ to apply to each qubit. The user can specify exactly how different errors combine $e_1e_2=e_3$. Each time a quantum gate is called, it is passed an error rate $p$ and an error of any kind is applied with probability $px$. Error $i$ is then applied with relative probability $s_i/\Sigma s_i$. The final line is the gate duration in arbitrary units. For discussion purposes, we include the internal representation of the error model below, which makes use of explicit relative probabilities rather than integers.
\begin{eqnarray}
&&2\nonumber\\
&&1.0\nonumber\\
&&6\nonumber\\
&&0.332~~0~~3\nonumber\\
&&0.057~~1~~0\nonumber\\
&&0.053~~2~~1\nonumber\\
&&0.115~~2~~2\nonumber\\
&&0.320~~3~~0\nonumber\\
&&0.123~~3~~2\nonumber\\
&&1
\end{eqnarray}

Error models are not only used to generate stochastic errors. Every time a gate is applied, all possible errors are generated and added to the list of errors on each qubit touched by the gate. The probability $pxs_i/\Sigma s_i$ is recorded in each error data structure. Each error is given a unique label. Multiple-qubit errors are represented by single-qubit errors on each qubit, each with the same label. If one executes a long sequence of unitary gates, the number of errors per qubit that need to be tracked will grow (linearly) without bound. Each unitary gate transforms all errors present on all of the qubits it touches. An $H$ gate will transform $X$ errors into $Z$ errors and vice versa. Multiple-qubit gates can create new propagated errors which will have the same label as the original and can combine or cancel multiple errors on a single qubit with the same label. For example, CNOT($q_1$, $q_2$, $p$ = 0.01) applied to qubits
\begin{eqnarray}
q_1&\rightarrow&(X, 0.00500, 0)\nonumber\\
   &\hookrightarrow&(Z, 0.00238, 1)
\end{eqnarray}
\begin{eqnarray}
q_2&\rightarrow&(Y, 0.00238, 1)
\end{eqnarray}
where ($A$, $p_{sr}$, $L$) represents the error type, scaled relative probability and label and the arrows represent a linked list, will result in
\begin{eqnarray}
q_1&\rightarrow&(X, 0.00500, 0)\nonumber\\
   &\hookrightarrow&(X, 0.00057, 3)\nonumber\\
   &\hookrightarrow&(Z, 0.00053, 4)\nonumber\\
   &\hookrightarrow&(Z, 0.00115, 5)\nonumber\\
   &\hookrightarrow&(Y, 0.00320, 6)\nonumber\\
   &\hookrightarrow&(Y, 0.00123, 7)
\end{eqnarray}
\begin{eqnarray}
\label{q2afterCNOT}
q_2&\rightarrow&(X, 0.00500, 0)\nonumber\\
   &\hookrightarrow&(Y, 0.00238, 1)\nonumber\\
   &\hookrightarrow&(Y, 0.00332, 2)\nonumber\\
   &\hookrightarrow&(X, 0.00053, 4)\nonumber\\
   &\hookrightarrow&(Z, 0.00115, 5)\nonumber\\
   &\hookrightarrow&(Z, 0.00123, 7)
\end{eqnarray}

\subsection{Sets of measurements}
\label{Sets of measurements}

Autotune currently supports only single-qubit measurements in the $X$ and $Z$ bases. $M_X$ applied to eq.~\ref{q2afterCNOT} will create a measurement
\begin{eqnarray}
m&\rightarrow&(Z, 0.00238, 1)\nonumber\\
 &\hookrightarrow&(Z, 0.00332, 2)\nonumber\\
 &\hookrightarrow&(Z, 0.00115, 5)\nonumber\\
 &\hookrightarrow&(Z, 0.00123, 7)
\end{eqnarray}
The $X$ errors and $X$ components of $Y$ errors have been removed. All errors will be removed from the qubit. To use the qubit again it must be explicitly initialized. If the qubit does not need to be used immediately after being measured, Autotune provides a dead gate that advances the qubit in time but does not generate or track any errors. This models incoherent evolution.

Every measurement is associated with either two sets or a single set and a boundary. A set of measurements has the property that the product of the measurement results (+1 or -1) indicates whether a chain of errors has ended nearby. In the standard surface code, sets contain consecutive pairs of syndrome qubit measurements. In a 3-D topological cluster state, sets contain the measurements on the faces of individual primal and dual cells. See Fig.~\ref{sets}. Sets must be specified by the user.

Sets can also be associated with boundaries. The bottom layer of sets in Fig.~\ref{sets}a is associated with the primal initial time boundary. In the second layer of sets, sets in the left row are associated with the left primal boundary, those in the right row are associated with the right primal boundary. We use the terminology primal/dual instead of rough/smooth as used in \cite{Fowl08} to ensure uniform terminology when discussing both the surface code and 3-D topological cluster states. Note that the middle row of sets in the second layer is not associated with any boundary. The association of sets with boundaries is currently manually user specified.

\subsection{Detection events}
\label{Detection events}

When all measurements in a set have been performed, further processing is triggered. A measurement may contain many errors. A set may contain many measurements. Autotune determines which errors with the same label appear an odd number of times. For each such label a detection event is generated. Fig.~\ref{syn_err}a contains an example of a single error leading to a pair of detection events. Detection events are stored in a hash table to enable one to quickly determine whether a detection event with a given label has already been generated. Pairs of detection events immediately trigger the creation of sticks, described in the following subsection.

\begin{figure}
\begin{center}
\resizebox{80mm}{!}{\includegraphics{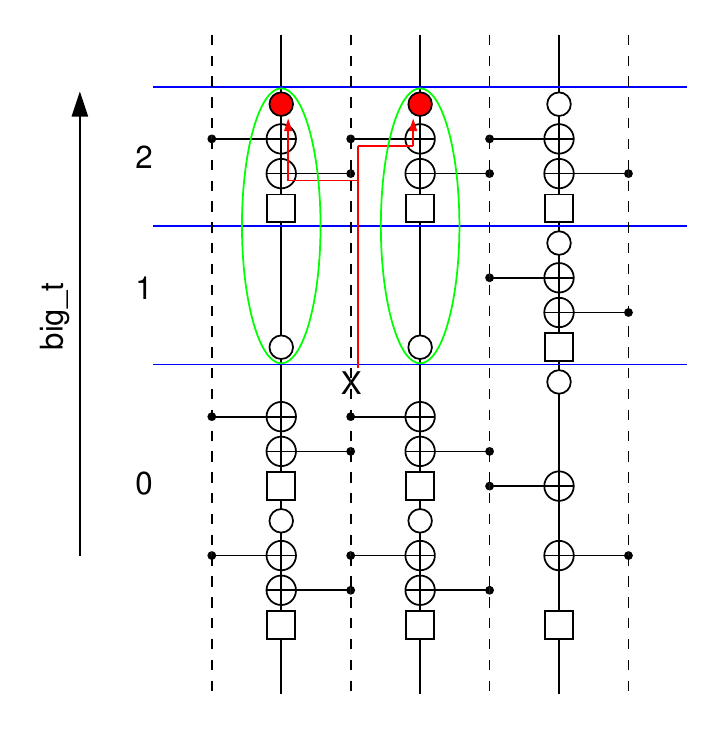}}
\end{center}
\caption{Example of a single error leading to two detection events two big\_t in the future. A detection event has a big\_t equal to that of its latest measurement. big\_t increments after the every stabilizers has been measured at least once (circles represent measurement, squares represent initialization).}\label{big_t}
\end{figure}

Errors near a boundary can lead to single detection events (Fig.~\ref{syn_err}b). Such single detection events must generate sticks leading to the nearby boundary. One must decide with care when to conclude that a detection event is unique and that no matching detection event will be generated in the future. To deal with this, we define a measure of error detection progress big\_t to increment only when every stabilizer of the code has been measured at least once. An example of an error leading to detection events two big\_t in the future is shown in Fig.~\ref{big_t}. This is the maximum delay possible. Detection events three big\_t in the past that are unique are guaranteed to remain so.

Detection events can be primal or dual depending whether they are associated with primal or dual sets which are in turn associated with primal or dual stabilizer measurements. It is a good idea to design stabilizer measurement circuits with the property that single errors do not lead to the creation of more than one pair of primal and one pair of dual detection events. This ensures that minimum weight perfect matching \cite{Edmo65a,Edmo65b} is well-suited to correcting errors generated during the execution of the circuits.

\subsection{Nests of balls and sticks}
\label{Nests}

When a set is processed to generate detection events, it is said to be finalized. At this point in time, a ball is generated and associated with the finalized set. This ball represents a location in space-time. A pair of detection events leads to the creation of a stick between its associated balls. A stick represents a potential connection between a pair of space-time locations arising from a single error. Many single errors can lead to the same stick \cite{Wang10}. The probability of a given stick, given a list of errors leading to it each with its own probability ${(e_i, p_i)}$, is therefore, to first order
\begin{equation}
p_{\rm stick} = \sum_i p_i \prod_{j\ne i} (1-p_j).
\label{pstick}
\end{equation}

We have a simple Blender based visualization tool for nests, used to generate Fig.~\ref{Manhattan_sc}, Fig.~\ref{Autotune_sc}, Fig.~\ref{Asymmetric_sc}, and Fig.~\ref{cluster}. A nest contains full tracking information of which gate led to which collection of errors and which of those errors led to which sticks. This is highly useful, however computationally cumbersome. Autotune does delete all errors, measurements, sets, detection events, balls and sticks when they are no longer required, which keeps total memory required finite, however it remains challenging to generate the nest fast enough to obtain good statistics in simulations, let alone keep pace with a real quantum computer. Note that nest generation is computationally efficient in the computer science sense, with constant memory and $O(n^2)$ time required to generate each layer of the nest on $n^2$ qubits, however mere efficiency is insufficient for practical purposes.

\subsection{Exploiting regularity}
\label{Lattices}

Motivated by the difficulty of rapidly generating nests, we first cut down the data stored therein to the minimum required by the minimum weight perfect matching algorithm. For every ball in a nest we create a dot, which again simply represents a space-time location. For every stick, we create a line. Lines connect the dots corresponding to the balls the stick connected, however they contain just one number, the weight $w=-\ln(p_{\rm stick})$.

The lattice is kept for much longer than the nest is as with low probability the matching algorithm can require all prior matching history to correctly match the latest data. A sufficiently long history must be kept to ensure the probability of requiring more is negligible. This is possible as the probability of requiring additional data in the past decreases exponentially with kept history size. The simplicity of the lattice keeps the memory required low.

The nests shown in the all have a great deal of regularity in their structure. This can be exploited to enable direct generation of lattices, avoiding the need to generate expensive nests on the fly. During a boot-up phase, Autotune analyzes each new stick and stores new unique sticks as an offset that contains only the geometric information associated with the stick --- just enough to create a line. Furthermore, each new ball is analyzed to determine if it corresponds to a new pattern of offsets. Such unique patterns are stored as blocks of offsets. Finally, each round of error detection is analyzed with structurally unique rounds stored as layers of blocks. All data is then stored in a recipe that contains all necessary information to rapidly create any part of the lattice.

Currently, Autotune is capable of analyzing either error detection circuitry that eventually leads to identical repeated layers or a finite number of rounds of error detection that can have any structure whatsoever. The former could easily be extended to a finite number of cyclically repeated layers, however it remains unclear whether one could avoid generating full nests when simulating probabilistic error detection in which all syndrome measurements can take a randomly variable amount of time.

The performance of Autotune has been described elsewhere \cite{Fowl12c}. Its complexity is optimal, requiring only $O(n^2)$ time to simulate and perform the necessary classical processing associated with each round of error detection on an array of $n^2$ qubits. Given constant computing power per unit area, all algorithms within Autotune can be parallelized to $O(1)$. This optimal parallelization is a direct consequence of the topological nature of the codes used and the fact that on average this implies that one only needs local information to correctly process a given space-time region of measurement results. The local runtime does not depend in any way on the global size of the computer.

\end{document}